\documentclass{article} % For LaTeX2e
\usepackage{iclr2027_conference,times}

\usepackage{amsmath,amsfonts,bm}

\def\eqref#1{equation~\ref{#1}}
\def\1{\bm{1}}

\DeclareMathAlphabet{\mathsfit}{\encodingdefault}{\sfdefault}{m}{sl}
\SetMathAlphabet{\mathsfit}{bold}{\encodingdefault}{\sfdefault}{bx}{n}

\newcommand{\R}{\mathbb{R}}

\usepackage{hyperref}
\usepackage{url}

\usepackage{amsmath}
\usepackage{amssymb}
\usepackage{amsthm}
\usepackage{bm}
\usepackage{graphicx}
\usepackage{booktabs}
\usepackage{xcolor}
\usepackage{enumerate}

\def\calD{{\mathcal{D}}}

\def\calK{{\mathcal{K}}}

\def\calT{{\mathcal{T}}}
\def\calU{{\mathcal{U}}}

\def\ctx{{\mathrm{ctx}}}
\def\disc{{\mathrm{disc}}}
\def\map{{\mathrm{map}}}
\usepackage{tikz}
\usetikzlibrary{calc}
\usetikzlibrary{arrows.meta}
\usetikzlibrary{positioning}

\usepackage{wrapfig}
\usepackage[colorinlistoftodos,prependcaption,textsize=tiny]{todonotes}

\DeclareMathOperator{\EX}{\mathbb{E}}

\newtheorem{thm}{Theorem}[section]
\newtheorem{cor}[thm]{Corollary}
\newtheorem{lem}[thm]{Lemma}
\newtheorem{prop}[thm]{Proposition}

\theoremstyle{definition}
\newtheorem{defn}[thm]{Definition}
\newtheorem{assum}[thm]{Assumption}

\title{Latent Twin Operator}

\author{Deepanshu Verma \\ Department of Mathematical and Statistical Sciences \\ Clemson University \\ \texttt{dverma@clemson.edu}
\\
  \And
  Riley Chen \\
  Department of Mathematics \\
  Emory University \\
  \texttt{yizhou.chen@emory.edu} \\
  \AND
  Matthias Chung \\
  Department of Mathematics \\
  Emory University \\
  \texttt{matthias.chung@emory.edu}
  }

\iclrfinalcopy % Uncomment for camera-ready version, but NOT for submission.

\begin{document}

\maketitle

\begin{abstract}
Surrogate models deployed on real physical systems rarely see data at fixed resolutions: sensor configurations vary across deployments and may evolve over time as sensing infrastructure changes. We introduce the Latent Twin Operator (LTO), a latent-space surrogate for time-evolving PDEs with a Convolutional Conditional Neural Process (ConvCNP)-style encoder and decoder that accepts a context set of $N$ sensor observations with arbitrary placement and can be queried at any resolution. A learned latent evolution map advances the encoded state directly between arbitrary time points, decoupling temporal evolution from the observation and query discretizations. We derive an explicit $\mathcal{O}(N^{-2/(3D)})$ rate for the context discretization error in spatial dimension $D$ under quasi-uniform refinement. We verify the predicted decay empirically on a 2D heat-equation benchmark. Across time-dependent PDE benchmarks, LTO achieves strong accuracy under one-step comparisons and transfers from native to coarser spatial resolutions with fixed parameters. On Navier--Stokes, its direct latent evolution further reduces error over longer prediction horizons relative to recursive evaluation.
\end{abstract}

\section{Introduction}
\label{sec:intro}

Realistic PDE workflows rarely operate at a single fixed resolution. In simulation, adaptive meshes refine where the dynamics demand it; in experiments and deployed systems, sensor locations and densities vary across settings and may change over time. Such settings call for learned surrogates that are not tied to the discretization on which they were trained, but can accept sparse, dense, or irregular observations and return predictions at whatever resolution is required. We refer to this ability to operate consistently as the input and output discretizations change as \emph{discretization invariance}.

To achieve this form of discretization invariance, we introduce the \emph{Latent Twin Operator} (LTO), an encode$\to$evolve$\to$decode surrogate in which observations are mapped to a latent state, advanced in time by a learned evolution map, and decoded at arbitrary query points. Its encoding and decoding employ parameter-free kernel-smoothing operations in the style of Convolutional Conditional Neural Processes \citep{gordon2020convolutional}, coupled to convolutional latent representations and dynamics (Definition~\ref{def:lto}).

Two standard operator-learning architectures impose stronger discretization requirements than the LTO. The Fourier Neural Operator (FNO; \citealp{li2020fourier}) is formulated on regular grids, since its core operation uses the FFT; nonuniform Fourier transforms or interpolation can relax this requirement, but require additional machinery outside the standard formulation. DeepONet \citep{lu2019deeponet} allows arbitrary output query locations through its trunk network, but its branch network is tied to a fixed set of input sensor locations. Therefore, in their standard forms, neither architecture directly accepts a variable-size, irregularly placed context set without resampling or modification. More generally, evaluating learned PDE surrogates on discretizations different from those used during training can introduce \emph{discretization mismatch error} \citep{gao2025discretization}.

The LTO builds directly on the Latent Twin framework \citep{chung2026latenttwins,chung2027physics}, which formalized the encode$\to$evolve$\to$decode Latent Twin formulation and established an approximation theory for latent-space surrogate models. LTO advances this framework through resolution-flexible encoding and decoding and derives a new explicit discretization-error bound, providing a theoretical guarantee of discretization invariance.

Among related latent operator methods, the closest in spirit is the Latent Neural Operator (LNO; \citealp{wang2024latent}). Like LTO, LNO maps a variable-size set of observations to a fixed-size latent representation and can decode at arbitrary query points. LNO achieves this flexibility through learned cross-attention over latent tokens, whereas LTO uses a spatially structured kernel-smoothing representation together with convolutional latent dynamics. The LTO construction therefore retains flexibility in the input and output discretizations while making the associated discretization error directly amenable to analysis. Other latent-space approaches, including the Latent Neural PDE Solver \citep{li2025latent} and the Physics-Informed Latent Neural Operator \citep{karumuri2026physics}, share elements of the encode$\to$evolve$\to$decode paradigm but differ in how the latent representation, evolution, and input--output maps are realized. We discuss these connections in more detail in Section~\ref{sec:related} and Appendix~\ref{app:related}.

A key advantage of the LTO's kernel-smoothing construction is its connection to the well-established theory of nonparametric regression and scattered-data approximation \citep{wasserman2006all,wendland2005scattered}. We use these tools to quantify how the approximation error depends on the number and placement of context points and derive an explicit convergence rate. This resolves the previously abstract spatial-discretization component of the general Latent Twin approximation theory \citep{chung2026latenttwins} and makes its dependence on context resolution explicit.

Our main contributions are:
\begin{itemize}
    \item We introduce the Latent Twin Operator, a resolution-flexible encode$\to$evolve$\to$decode surrogate that accepts context sets of varying size and placement and produces predictions at arbitrary query resolutions. For its SetConv-based realization, we prove encoder stability with a Lipschitz constant independent of context-set size and placement.

    \item We derive an explicit spatial discretization-error bound using scattered-data approximation theory, $\epsilon_{disc}(N)=O(N^{-2/(3D)})$, for $N$ context points in spatial dimension $D$. Combined with the general Latent Twin approximation theory, this yields a discretization-invariance guarantee under quasi-uniform refinement.
    
    \item We verify the predicted discretization decay on a 2D heat-equation benchmark and evaluate the learned LTO under matched protocols on Burgers, Navier--Stokes, and Darcy flow. On the time-dependent benchmarks, LTO transfers from the native training resolution to coarser spatial resolutions with fixed model parameters and achieves lower matched one-step errors than FNO on Burgers and than FNO and LNO on Navier--Stokes. On Navier--Stokes, direct latent evolution further reduces long-horizon error relative to recursive evaluation.

\end{itemize}

Section~\ref{sec:lto} develops the Latent Twin Operator and its discretization-invariance theory. Section~\ref{sec:related} places the LTO in the broader operator-learning landscape. Section~\ref{sec:experiments} evaluates its discretization behavior across four PDE benchmarks -- the 2D heat equation, Burgers, 2D Navier--Stokes, and Darcy flow -- under a common train-once, evaluate-across-resolutions protocol. Section~\ref{sec:discussion} examines the scope and limitations of the regularity assumptions underlying the theory. Additional technical details, proofs, and extended comparisons with related methods are provided in the Appendix~\ref{sec:App}.

\section{The Latent Twin Operator}
\label{sec:lto}

We consider time-evolving states $u:\calT\to\calU$ defined on a bounded spatial domain $\Omega\subset\R^D$, where $\calT=[0,T]$ and $\calU$ is a normed vector space of functions from $\Omega$ to $\R^p$. At any time $t$, the state $u(t)$ is observed $N$ at context locations $c_{\ctx}=\{c_1,\ldots,c_N\}\subset\Omega$. For any state $u_t=u(t)$, we denote the corresponding observed context values by $ u_t(c_{\ctx}) := \{u_t(c_j)\}_{j=1}^N$. The state may be evaluated at an arbitrary set of $M$ query locations $ c_{\mathrm q}=\{q_1,\ldots,q_M\}\subset\Omega,$ with corresponding values $ u_t(c_{\mathrm q}) := \{u_t(q_j)\}_{j=1}^M.$
Different observations of the same underlying state may use context and
query sets of different size and placement, precisely the multi-resolution
setting motivated in Section~\ref{sec:intro}.

\begin{defn}[Latent Twin Operator]
\label{def:lto}
Let $\Phi(\cdot\,;s,u_s):\calT\to\calU$ denote the solution of $ \dot u=\calD(u)$ with $ u(s)=u_s$. The \emph{Latent Twin Operator} (LTO) is the surrogate 
\begin{equation}
    \Psi^{s,c_{\ctx},u_s}(t,c_{\mathrm q}) := d\!\left( m_{s\to t}\!\left( e\!\left(u_s(c_{\ctx}),c_{\ctx}\right) \right), c_{\mathrm q} \right).
\label{eq:lto}
\end{equation}
Here, $e$ maps the observed context data to a latent state, $m_{s\to t}:\mathcal Z\to\mathcal Z$ evolves this state from time $s$ to time $t$, and $d$ maps the evolved latent state to values at the query locations $c_{\mathrm q}$.
\end{defn}

The encoder $e$, latent evolution map $m_{s\to t}$, and decoder $d$ are trainable; we suppress their learnable parameter dependence throughout for notational simplicity, writing it explicitly as $\Psi_\theta$ below where the discretization bound requires it. We next specify the encoder $e$, latent evolution map $m_{s\to t}$, and decoder $d$. The encoder and decoder are chosen to accommodate arbitrary context and query discretizations, while $m_{s\to t}$ evolves the resulting representation entirely in latent space.

\paragraph{Resolution-independent encoding.}
To accommodate context sets of arbitrary size and placement, we construct the encoder using the Convolutional Conditional Neural Process (ConvCNP) framework \citep{gordon2020convolutional}, extending Conditional Neural Processes~\citep{garnelo2018conditional}. Given the context data $\big(c_{\ctx},u_s(c_{\ctx})\big)$, a set convolution (SetConv) maps this unordered, variable-size set to a field on a fixed spatial grid. Each observation contributes to nearby locations through a distance-based kernel, making the construction invariant to the ordering of the context points while retaining their spatial locations.

We use the Gaussian kernel $\kappa_\ell(x,y) = \exp\!\left( -\frac{d_\Omega(x,y)^2}{2\ell^2} \right)$, where $d_\Omega$ denotes the distance on $\Omega$ and $\ell>0$ determines the kernel length scale. For a generic spatial location $x\in\Omega$, define
\begin{equation}
    \hat u_{c_{\ctx}}(x;u_s) := \frac{\nu_{c_{\ctx}}(x;u_s)} {\rho_{c_{\ctx}}(x)},
    \label{eq:normalized-signal-channel} \;\;\; \text{where}
\end{equation}
\begin{equation}
    \rho_{c_{\ctx}}(x) := \sum_{j=1}^N \kappa_\ell(x,c_j)
    \qquad \text{and}\qquad
    \nu_{c_{\ctx}}(x;u_s) := \sum_{j=1}^N u_s(c_j)\kappa_\ell(x,c_j).
    \label{eq:density-signal-channel}
\end{equation}
The quantity $\hat u_{c_{\ctx}}$ is the Nadaraya--Watson kernel estimate of $u_s$ \citep{nadaraya1964estimating,watson1964smooth}; $\rho_{c_{\ctx}}$ measures the local concentration of context points, while $\nu_{c_{\ctx}}$ is the corresponding kernel-weighted sum of state values.

We evaluate $\rho_{c_{\ctx}}$ and $\hat u_{c_{\ctx}}$ on a fixed spatial grid and stack them into a $(1+p)$-channel field. The encoder $e$ maps this gridded SetConv representation to
\[
    z_s = e\!\left( \rho_{c_{\ctx}}, \hat u_{c_{\ctx}}(\,\cdot\,;u_s) \right) \in\mathcal Z:=\R^{H\times W\times k}.
\]

\paragraph{Latent evolution.}
The latent evolution map  $
m_{s\to t}:\mathcal Z\to\mathcal Z
$
advances the encoded state directly from time $s$ to time $t$. Since $m_{s\to t}$ acts entirely on the fixed latent representation, its input and output are independent of the context and query discretizations.

\paragraph{Resolution-independent decoding.}
Given the evolved latent state $z_t$, the decoder $d$ maps the latent representation back to the physical state and evaluates it at the query locations $c_{\mathrm q}$. Its final evaluation uses kernel-weighted interpolation on the spatial grid, so the number and placement of query points may vary independently of both the context locations and the internal grid.

\begin{figure}
    \centering
    \includegraphics[width=\linewidth]{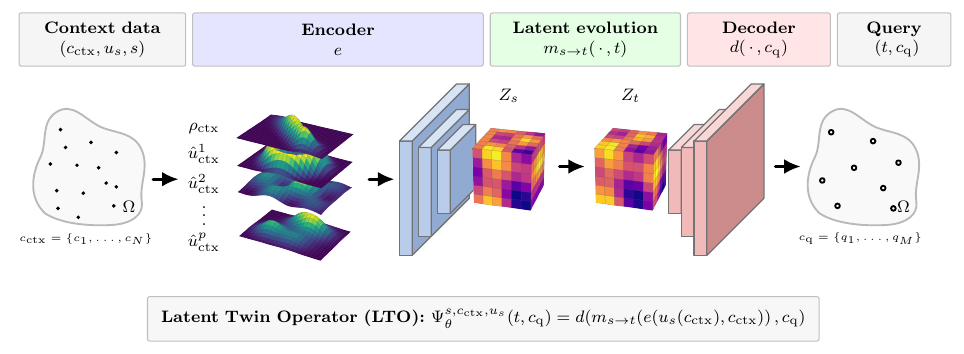}
    \caption{Latent Twin Operator. Irregular context observations are mapped by SetConv to a fixed-grid representation, encoded into latent space, evolved from $Z_s$ to $Z_t$ by $m_{s\to t}$, and decoded at arbitrary query locations.}
    \label{fig:lto}
\end{figure}

\paragraph{Connection to the Latent Twin \citep{chung2026latenttwins} approximation theory.}
The LTO extends the general Latent Twin framework \citep{chung2026latenttwins} to settings with variable context and query discretizations. While the original framework develops its approximation theory under a fixed discretization and treats the encoder, latent evolution map, and decoder abstractly, here we construct these components in a resolution-flexible form and use that theory as the basis for an explicit discretization analysis.

For each $N$, first fix a context discretization $c_{\ctx}^{(N)}$ and define 
$P_N:\mathcal U\to\R^{Np}$ with $P_Nu=u(c_{\ctx}^{(N)})$ as the corresponding sampling operator. Following \citet[Thm.~2]{chung2026latenttwins}, the query set is taken equal to the context set, $c_{\mathrm q}=c_{\ctx}^{(N)}$, throughout this section. Let $R_N:\R^{Np}\to\mathcal U$ be a lifting operator such that the pair $(P_N,R_N)$ is consistent and stable in the sense required by \citet[Thm.~2]{chung2026latenttwins}. The general Latent Twin theory applies to any such consistent and stable pair
$(P_N,R_N)$ for which
\[
    \epsilon_{\disc}(N) := \sup_{u\in\mathcal K} \|R_N(P_Nu)-u\|_{\mathcal U} \longrightarrow 0,
\]
where $\mathcal K\subset\mathcal U$ is a compact set of admissible states.

We denote by
\begin{equation} \label{eq:lifted-lto}
    \widetilde\Psi_{\theta,N}(t;s,u_s) := R_N\Big(\Psi_\theta^{\,s,c_{\ctx}^{(N)},u_s}\big(t,c_{\ctx}^{(N)}\big)\Big)
\end{equation}
the corresponding lifted LTO, where $\Psi_\theta$ is as in Definition.~\ref{def:lto} with its parameter dependence made explicit.
Under the additional assumptions of \citet[Thm.~2]{chung2026latenttwins}, the corresponding lifted Latent Twin satisfies
\begin{equation} \label{eq:base-bound}
    \sup_{(s,t,u_s)\in[0,T]^2\times\mathcal K} \left\| \widetilde\Psi_{\theta,N}(t;s,u_s) - \Phi(t;s,u_s) \right\|_{\mathcal U}
    \le
    C\left(\epsilon_{\disc}(N) + \bigl(1+\mathrm e^{L_{\mathcal K}T}\bigr) \epsilon_{\mathrm{ae}}^{(N)} + \,L_{d,N}\,\epsilon_{\map}^{(N)}\right).
    % C\,\epsilon_{\disc}(N) + C\bigl(1+\mathrm e^{L_{\mathcal K}T}\bigr) \epsilon_{\mathrm{ae}}^{(N)} + C\,L_{d,N}\,\epsilon_{\map}^{(N)}.
\end{equation}

Here $\epsilon_{\mathrm{ae}}^{(N)}$ is the autoencoder approximation error and $\epsilon_{\map}^{(N)}$ is the latent-evolution error. The quantity $L_{d,N}$ is the decoder stability constant from \citet{chung2026latenttwins}, $L_{\mathcal K}$ controls the Lipschitz growth of the underlying dynamics on $\mathcal K$, and $C>0$ is a constant determined by the assumptions of the theorem.
For each $N$, the bound applies to any fixed admissible pair $(P_N,R_N)$ satisfying the assumptions of \citet[Thm.~2]{chung2026latenttwins}, regardless of the particular discretization and reconstruction operators used.

For the LTO, we take $P_N$ to be sampling at the context locations and instantiate $R_N$ by the normalized SetConv kernel reconstruction,
\[
    R_N(P_N u)(x)
    :=
    \frac{\sum_{j=1}^N u(c_j)\kappa_\ell(x,c_j)}
         {\sum_{j=1}^N \kappa_\ell(x,c_j)}.
\]
This is precisely the normalized kernel estimate \eqref{eq:normalized-signal-channel} used in the SetConv representation. Thus, the abstract discretization term $\epsilon_{\disc}(N)$ in \eqref{eq:base-bound} becomes a concrete scattered-data approximation error. We first establish stability of this reconstruction and then derive an explicit bound on $\epsilon_{\disc}(N)$ in terms of the number and placement of the context points.
\paragraph{Encoder stability.}
Before deriving the discretization rate, we first establish that the normalized SetConv reconstruction is stable with respect to perturbations of the observed state. For a fixed context set $c_{\ctx}$, the Nadaraya--Watson reconstruction satisfies a Lipschitz bound whose constant does not depend on the number or placement of the context points. Hence the stability of the SetConv preprocessing is uniform across admissible context discretizations. The precise statement and proof are given in Lemma~\ref{lem:encoder-stability} and Corollary~\ref{cor:Le} in Appendix~\ref{app:proofs}.

\paragraph{Explicit discretization error.}
For the LTO, the sampling operator $P_N$ records the values of $u$ at the context locations, while the reconstruction operator $R_N$ is the normalized SetConv kernel reconstruction. Hence $R_N(P_Nu) = \hat u_{c_{\ctx}}(\,\cdot\,;u)$, and therefore
\[
    \epsilon_{\disc}(N) = \sup_{u\in\mathcal K} \left\| \hat u_{c_{\ctx}}(\,\cdot\,;u)-u \right\|_{\mathcal U}.
\]
Thus $\epsilon_{\disc}(N)$ is the error of kernel smoothing on scattered data. Its value depends on both the number of context points and their placement. To obtain an explicit rate, we impose the following regularity condition on the admissible states and context sets.

\begin{assum}[Regularity]
\label{assum:reg}
Assume that (a) $ \mathcal K\subset C^2(\Omega;\mathbb R^p)$ satisfies $\sup_{u\in\mathcal K}\sup_{x\in\Omega}\|\mathrm{D}^2u(x)\|\le M$, (b) $\Omega$ is periodic, and (c) that each context set
$c_{\ctx}=\{c_1,\ldots,c_N\}$ admits a partition
$V_1,\ldots,V_N$ of $\Omega$ such that
\[
c_j\in V_j,\qquad
\operatorname{diam}(V_j)\le 2\delta(c_{\ctx}),\qquad
|V_j|=\frac{|\Omega|}{N}
\]
where
$
\delta(c_{\ctx})
:=
\sup_{x\in\Omega}\min_{1\le j\le N} d_\Omega(x,c_j)
$
is the fill distance.
\end{assum}

Condition~(a) controls the local curvature of the admissible states, while condition~(c) prevents the context points from becoming arbitrarily clustered. The equal-volume partition is slightly stronger than standard quasi-uniformity assumptions in scattered-data approximation \citep{wendland2005scattered}, but is used directly in the quadrature argument of Proposition~\ref{prop:quadrature}.

Under Assumption~\ref{assum:reg}, we decompose the SetConv reconstruction error into a kernel-smoothing component and a finite-context component by introducing the continuous analogue
\[
    \bar u_\ell(x) := \frac{\int_\Omega u(y)\kappa_\ell(x,y)\,\mathrm{d}y} {\int_\Omega \kappa_\ell(x,y)\,\mathrm{d}y}.
\]
Under the periodicity assumption, the denominator is independent of $x$ (see Lemma~\ref{lem:kbar-constant}); we denote it by $ \bar\kappa := \int_\Omega \kappa_\ell(x,y)\,\mathrm{d}y.$ The quantity $\bar u_\ell$ is obtained by replacing the finite kernel-weighted sums in $\hat u_{c_{\ctx}}$ by their corresponding integrals. It therefore provides the decomposition
\[
    \|\hat u_{c_{\ctx}}-u\|_{\mathcal U} \le \|\hat u_{c_{\ctx}}-\bar u_\ell\|_{\mathcal U} + \|\bar u_\ell-u\|_{\mathcal U},
\]
where the second term is the kernel-smoothing bias and the first is the finite-context quadrature error.

The smoothing error follows from a second-order Taylor expansion. Because the Gaussian kernel is symmetric about $x$, the first-order term vanishes after integration, leaving a contribution controlled by the Hessian of $u$. Under Assumption~\ref{assum:reg}, this gives
\begin{equation} \label{eq:bias-bound}
    \sup_{u\in\mathcal K}\sup_{x\in\Omega} \|\bar u_\ell(x)-u(x)\| \le \tfrac{1}{2}DM\ell^2.
\end{equation}
The proof is given in Appendix~\ref{app:proofs};
see also \citet[Ch.~5]{wasserman2006all} for the corresponding classical kernel-smoothing argument.

It remains to control the finite-context error $\|\hat u_{c_{\ctx}}-\bar u_\ell\|_{\mathcal U}$. Under Assumption~\ref{assum:reg}, a scattered-data quadrature argument (Proposition~\ref{prop:quadrature}, Appendix~\ref{app:proofs}) shows that this error is of order $\delta(c_{\ctx})/\ell$. Together with the $O(\ell^2)$ smoothing bias from~\eqref{eq:bias-bound}, this yields the trade-off quantified in the following theorem.

\begin{thm}[Explicit discretization error for the SetConv instantiation] \label{thm:eps-disc}
Under Assumption~\ref{assum:reg}, for any bandwidth $\ell>0$,
\[
    \epsilon_{\disc}(N) \le \tfrac12 DM\,\ell^2 + C_{\mathcal K}\frac{\delta(c_{\ctx})}{\ell},
\]
where $C_{\mathcal K}>0$ depends only on $D$, $\Omega$, $M$, and $\sup_{u\in\mathcal K}\|u\|_\infty$.
\end{thm}

\begin{cor}[Optimized discretization rate]
\label{cor:optimized-disc}
Choosing $ \ell^\star = \left( \frac{C_{\mathcal K}\,\delta(c_{\ctx})}{DM} \right)^{1/3} $ in Theorem~\ref{thm:eps-disc} gives
\[
    \epsilon_{\disc}(N) \le C\,\delta(c_{\ctx})^{2/3},
\]
for a constant $C>0$ independent of the context resolution.
\end{cor}

\begin{cor}[Discretization invariance]
\label{cor:disc-invariance}
Assume that the context sets satisfy $ \delta(c_{\ctx})=\Theta(N^{-1/D})$. With the bandwidth choice $\ell=\ell^\star$ from Corollary~\ref{cor:optimized-disc}, substituting the resulting discretization rate into the Latent Twin Operator error bound~\eqref{eq:base-bound} gives
\[
    \sup_{(s,t,u_s)\in[0,T]^2\times\mathcal K} \left\| \widetilde\Psi_{\theta,N}(t;s,u_s) - \Phi(t;s,u_s) \right\|_{\mathcal U} 
    \le
    O\!\left(N^{-2/(3D)}\right) + C\bigl(1+\mathrm e^{L_{\mathcal K}T}\bigr) \epsilon_{\mathrm{ae}}^{(N)} + C L_{d,N}\epsilon_{\map}^{(N)}.
\]
\end{cor}
For i.i.d.\ uniformly sampled context points, the fill-distance estimate of \citet{penrose2003random} yields the corresponding high-probability rate
\[
    \epsilon_{\disc}(N) = O\!\left( \left(\frac{\log N}{N}\right)^{2/(3D)} \right).
\]

\section{Related Work}
\label{sec:related}

The Latent Twin Operator sits at the intersection of four lines of work: neural operators, implicit neural representations, conditional neural processes, and the latent-space encode$\to$evolve$\to$decode architectures it most closely resembles. We summarize each briefly here, with an extended discussion in Appendix~\ref{app:related}.

% Operator learning formalizes learning a PDE's solution operator directly \citep{kovachki2023neural,boulle2024mathematical}; DeepONet \citep{lu2019deeponet} and the Fourier Neural Operator (FNO; \citealp{li2020fourier}) are its two standard instantiations, the latter requiring a regular grid via the FFT. Our work has no grid dependency, it builds on Conditional Neural Processes. Conditional Neural Processes and their convolutional variant, ConvCNP \citep{garnelo2018conditional,gordon2020convolutional}, provide the kernel-weighted SetConv embedding underlies our encoder and decoder (Section~\ref{sec:lto}). Implicit representations such as SIREN \citep{sitzmann2020implicit} are discretization-agnostic per field but do not by themselves give an encode$\to$evolve$\to$decode surrogate for a family of states.
Operator learning formalizes learning a PDE's solution operator directly \citep{kovachki2023neural,boulle2024mathematical}; DeepONet \citep{lu2019deeponet} and the Fourier Neural Operator (FNO; \citealp{li2020fourier}) are its two standard instantiations, the latter requiring a regular grid via the FFT. Our encoder has no such grid dependency: it builds instead on Conditional Neural Processes and their convolutional variant, ConvCNP \citep{garnelo2018conditional,gordon2020convolutional}, whose kernel-weighted SetConv embedding underlies our encoder and decoder (Section~\ref{sec:lto}).

Closest to our method is the Latent Neural Operator (LNO; \citealp{wang2024latent}): its Physics-Cross-Attention likewise maps an arbitrary input set to a fixed-size latent state with input and output locations decoupled, but via learned cross-attention over an unstructured token set, rather than the parameter-free, spatially-structured kernel smoothing we use. 
A wider family of attention-based architectures tackles the same irregular-input problem by compressing to a smaller token set or by applying efficient attention over the full one (Appendix~\ref{app:related} surveys this family in detail); we compare against LNO specifically because it is architecturally closest to our construction, not because it is uniquely capable of cross-resolution generalization -- ONO \citep{xiao2023improved} reports the same zero-shot Darcy transfer.
This is what lets our latent evolution operator be a plain CNN and what admits the classical error analysis behind Theorem~\ref{thm:eps-disc}. No analogous discretization-error rate has been established for irregular-input encoders generally. The closest theoretical results, \citet{calvello2025continuum} for attention-based operators and \citet{zhang2025discretization} for a DeepONet-style branch-trunk construction, give universal approximation but not an explicit rate in the number of input points (Appendix~\ref{app:related} expands this comparison). The Latent Neural PDE Solver \citep{li2025latent} and Physics-Informed Latent Neural Operator \citep{karumuri2026physics} follow the same encode$\to$evolve$\to$decode shape without a comparable guarantee; and prior work, Latent Twins \citep{chung2026latenttwins,chung2027physics}, proves the approximation bound~\eqref{eq:base-bound} this paper builds on but leaves its discretization-error term $\epsilon_{\disc}(N)$ abstract. \citet{gao2025discretization} and \citet{lanthaler2024discretization} show, empirically and theoretically respectively, that FNO-style resolution-independence claims do not automatically hold.

\section{Experiments}
\label{sec:experiments}

We evaluate the LTO along four complementary axes: (i) whether the discretization behavior predicted by Theorem~\ref{thm:eps-disc} is observed empirically, (ii) one-step predictive accuracy, (iii) long-horizon behavior under direct and recursive temporal evolution, and (iv) transfer across spatial resolutions with fixed model parameters. Section~\ref{sec:exp-heat} isolates the discretization effect on a smooth 2D heat equation. Sections~\ref{sec:exp-burgers} and \ref{sec:exp-ns2d} evaluate the full learned model on Burgers' equation and 2D Navier--Stokes, respectively, while Section~\ref{sec:exp-ns2d} additionally examines long-horizon evolution. Section~\ref{sec:exp-darcy} considers Darcy flow as a steady-state resolution-transfer benchmark.

\paragraph{Setup.} Training data consists of trajectories $u$ observed
at a fixed set of native time steps.
The training objective is
\begin{equation}\label{eq:total-loss}
\mathcal{L}(\theta) = \lambda_{\mathrm{evol}}\underbrace{\EX\Big[\big\|\Psi_\theta^{s,c_{\ctx},u_s}(t,c_q) - u_t(c_q)\big\|_2^2\Big]}_{\mathcal{L}_{\mathrm{evol}}(\theta)}
+ \lambda_{\mathrm{recon}}\underbrace{\EX\Big[\big\|d_\theta\big(e_\theta(c_{\ctx};u_s),c_q\big) - u_s(c_q)\big\|_2^2\Big]}_{\mathcal{L}_{\mathrm{recon}}(\theta)}\,,
\end{equation}
where $\theta$ are the trainable parameters suppressed in
Section~\ref{sec:lto}, and $\mathcal{L}_{\mathrm{evol}}$,
$\mathcal{L}_{\mathrm{recon}}$ are the evolution and
reconstruction (autoencoder) losses, respectively.
Here, the expectation is taken over $n$ training examples
$\{(u_s^i,s^i),(u_t^i,t^i)\}_{i=1}^n$, each consisting of a trajectory
$u^i$ and a source-target time pair $(s^i,t^i)$ drawn from that
trajectory's native time discretization. The values of
$\lambda_{\mathrm{evol}},\lambda_{\mathrm{recon}}$ are stated for each benchmark below.
All error numbers are
relative $L^2$, $\|\Psi_\theta-u\|_2/\|u\|_2$ (dropping $\Psi_\theta$'s
arguments here, since they vary by evaluation setting). 
We additionally apply context-size augmentation% (denoted \textit{ctxaug})
, where during training we resample the context
fraction $c_{\mathrm{frac}} \in [c_{\min},c_{\max}]$ uniformly at random
on every step, holding the query and target at the full native
resolution throughout, and sample only the encoder's context set. Each benchmark's own $[c_{\min},c_{\max}]$ is stated in its
subsection below.

We use the standard time-evolving PDEBench~\citep{takamoto2022pdebench}
benchmarks, Burgers' equation and Navier--Stokes (Sections~\ref{sec:exp-burgers}
and~\ref{sec:exp-ns2d}). All models are trained from a single source state $u_s$ to the
corresponding target state $u_t$, at a source-target time pair $(s,t)$
drawn as above from each
trajectory's native time discretization. One-step accuracy is evaluated
under this same protocol; long-horizon behavior under direct and
recursive evaluation is examined separately, as detailed in the
relevant subsections below.

\subsection{2D Heat Equation: Resolution Sweep}
\label{sec:exp-heat}
We test discretization invariance (Corollary~\ref{cor:disc-invariance}) on a
2D periodic heat equation. 
\begin{wrapfigure}{r}{0.4\textwidth}
\vspace*{-3ex}
\centering
\includegraphics[width=\linewidth]{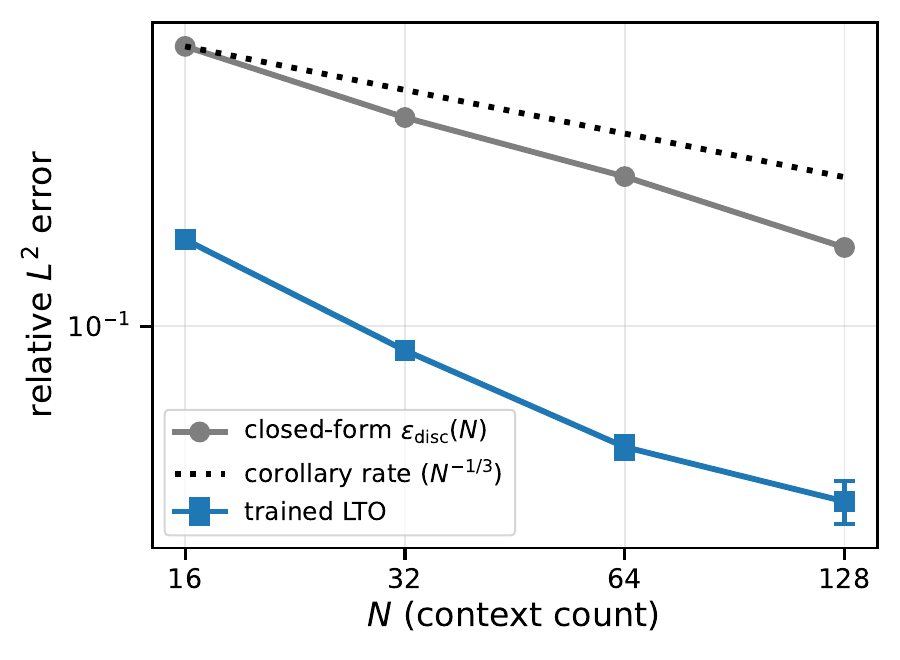}
\vspace*{-5.5ex}
  \caption{Closed-form $\epsilon_{\disc}(N)$ and trained LTO evaluated across context points $N$ without retraining, against Corollary~\ref{cor:disc-invariance}'s predicted rate.}
  \label{fig:heat-disc-vs-trained}
  \vspace{-22pt}
\end{wrapfigure}
The ground-truth field is a sum of four Gaussian
kernels (standard
deviations $0.12$--$0.22$) on the periodic 2D domain, evolved forward
exactly via an FFT-based spectral solve. Both kernels and their heat-flow evolution are smooth by construction, so Assumption~\ref{assum:reg} holds.

We test two related quantities against the context count $N$. The first is
$\epsilon_{\disc}(N)$ itself, computed directly from the closed-form
Nadaraya--Watson reconstruction \eqref{eq:normalized-signal-channel}, no
training involved.

The second is a single trained LTO's own forecasting error: given $N$ context points sampled from $u_s$, the network encodes them, evolves the latent state to a later time $t$, decodes a prediction, and we compare against the true field $u_t$ at that time. The network is trained once, with the context count resampled at random on every batch, and evaluated at every $N$ without retraining; the query count stays fixed throughout. We average this test over 5 runs. Both errors decrease at a rate consistent with the predicted slope, with a fitted log-log slope close to the
rate in Corollary~\ref{cor:disc-invariance}. 

\subsection{\texorpdfstring{Burgers' equation}{Burgers' equation}}
\label{sec:exp-burgers}

We evaluate LTO on PDEBench's Burgers equation at $\nu=0.001$, where solutions develop steep, near-shock fronts and therefore provide a substantially more demanding test of spatial resolution than the smooth heat-equation setting. We train LTO with $\lambda_{\mathrm{evol}}=\lambda_{\mathrm{recon}}=1$ and context-size augmentation $c_{\mathrm{frac}}\in[0.05,1.0]$, and compare against a self-trained FNO baseline under a matched training protocol. Both models are evaluated across resolutions from $64$ to $2048$ points without retraining, including the above-native $2048$-point interpolation regime. Table~\ref{tab:restransfer-burgers} summarizes the results.

LTO achieves lower one-step prediction error than FNO at every tested resolution, from $64$ to $2048$ points, including the above-native interpolated resolution. This advantage is therefore not confined to the training resolution or its immediate neighborhood. LTO also requires substantially less training time: approximately one quarter of FNO's wall-clock time, while using a comparable number of parameters ($29{,}395$ versus $23{,}937$). On this benchmark, LTO therefore combines stronger cross-resolution accuracy with lower training cost.

\begin{table}[t]
\centering
\caption{Burgers resolution transfer, one-shot relative $L^2$ (a single
evolution step). Resolutions at or below the native
resolution (1024) use real subsampled data; above-native points are
synthetic interpolation.}
\label{tab:restransfer-burgers}
\begin{tabular}{lccccccc}
\toprule
Method & Params & 64 & 128 & 256 & 512 & 1024 (native) & 2048 \\
\midrule
LTO (ours) & 29,395 & \textbf{0.078} & \textbf{0.063} & \textbf{0.060} & \textbf{0.056} & \textbf{0.057} & \textbf{0.066} \\
FNO        & 23,937 & 0.097 & 0.080 & 0.076 & 0.075 & 0.076 & 0.077 \\
\bottomrule
\end{tabular}
\end{table}
 
\begin{table}[t]
\centering
\caption{Navier Stokes (2D) resolution transfer, one-shot relative $L^2$,
averaged over six data points. FNO's fixed spectral mode truncation cannot run below
$32\times32$ -- an architectural constraint, not a missing measurement.}
\label{tab:restransfer-ns2d}
\begin{tabular}{lccccc}
\toprule
Method & Params & 16$\times$16 & 32$\times$32 & 64$\times$64 (native) & 128$\times$128 \\
\midrule
LTO (ours) & 1,126,211 & \textbf{0.154} & \textbf{0.147} & \textbf{0.146} & 6.156 \\
FNO        & 1,124,193 & --             & 0.213          & 0.177          & 1.608 \\
LNO        & 1,146,945 & 0.261          & 0.253          & 0.251          & \textbf{1.339} \\
\bottomrule
\end{tabular}
\end{table}

\subsection{Navier--Stokes (2D)}
\label{sec:exp-ns2d}
Two-dimensional Navier--Stokes is a substantially more challenging problem
than Burgers. We evaluate on PDEBench's 2D Navier--Stokes forward benchmark, trained
at resolution $64\times64$ to match LNO's own reported setting
\citep{wang2024latent}, and compare against self-trained FNO and LNO
baselines with parameter count matched to LTO's.
 
We train LTO with $\lambda_{\mathrm{evol}}=\lambda_{\mathrm{recon}}=1$,
context-size augmentation with $c_{\mathrm{frac}} \in [0.05, 4.0]$, and
the pushforward trick \citep{brandstetter2022message}: for $0.15$ of
training steps, the model unrolls autoregressively under no gradient
before one supervised step against the true target, exposing training
to the compounding error it otherwise meets only at inference.

Table~\ref{tab:restransfer-ns2d} reports one-step relative $L^2$ error across resolutions. At the sub-native and native resolutions ($16\times16$, $32\times32$, $64\times64$), LTO has the lowest error and remains nearly insensitive to coarsening; above native, at $128\times128$, all three models deteriorate, with LTO showing the largest increase. Thus, on this problem, the resolution flexibility of the representation translates into robust transfer to coarser discretizations, but not into reliable extrapolation to spatial
resolution finer than that used for training.

To assess long-horizon behavior, we recursively apply each model for
$k=1,\dots,10$ one-step increments. For LTO, we additionally evaluate
each target time directly from the source state through $m_{s\to t}$.
Table~\ref{tab:ns2d-long-horizon} shows a different ordering than
Table~\ref{tab:restransfer-ns2d}'s one-step comparison, since the
latter averages over six context start points while the former fixes
context at a particular time: FNO and LNO have lower error at $k=1$ but grow
faster under recursive evolution, and by $k=5$ and $k=10$ both LTO
evaluations are lower, with direct evaluation giving the lowest
ten-step average and confirming that avoiding repeated re-encoding
limits error growth.

\begin{table}[t]
\centering
\caption{Navier--Stokes long-horizon relative $L^2$ at the native $64\times64$ resolution. All methods start from the same source state at time $s$. ``Evolve'' recursively feeds each model's prediction back as the next source state for $k$ steps, whereas ``direct'' evaluates $m_{s\to t}$ from the original encoded source state without intermediate re-encoding.}
\label{tab:ns2d-long-horizon}
\begin{tabular}{lcccc}
\toprule
Method & $k=1$ & $k=5$ & $k=10$ & Avg.\ (10 steps) \\
\midrule
LTO, evolve (ours)      & 0.096 & 0.172 & 0.378 & 0.209 \\
LTO, direct eval (ours)  & 0.096 & \textbf{0.146} & \textbf{0.317} & \textbf{0.177} \\
FNO, evolve             & \textbf{0.055} & 0.194 & 0.705 & 0.294 \\
LNO, evolve             & 0.067 & 0.295 & 1.197 & 0.462 \\
\bottomrule
\end{tabular}
\vspace*{-0.75ex}
\end{table}

\subsection{Darcy Flow}
\label{sec:exp-darcy}

Darcy flow is a steady-state boundary-value problem mapping a piecewise-constant permeability field to the corresponding pressure field under a second-order elliptic PDE~\citep{li2020fourier}. Unlike the preceding benchmarks, it has no temporal evolution, so we include it primarily for a matched comparison with LNO. Following \citet{wang2024latent}, we train at $211\times211$ and evaluate without retraining at resolutions $85$, $106$, $141$, $211$, and $421$, using $\lambda_{\mathrm{recon}}=1$ and $c_{\mathrm{frac}}\in[0.05,1.0]$.

Because the Gaussian SetConv kernel of Section~\ref{sec:lto} imposes a strong locality bias, we also evaluate an attention-based variant with learned context weights; its definition is given in Appendix~\ref{app:darcy-native}. This variant is empirical and is not covered by the SetConv discretization theory.

Table~\ref{tab:darcy-results} reports relative $L^2$ errors and parameter counts. The Gaussian LTO trails LNO by $11.80\times$ on average, while the attention variant reduces this ratio to $4.86\times$, substantially narrowing the gap although remaining less accurate at every resolution. Appendix~\ref{app:darcy-native} also reports comparison with LNO's original $762{,}113$-parameter configuration.

\begin{table}[t]
\centering
\caption{Darcy flow: relative $L^2$ error across evaluation
resolutions, with LNO's parameter count matched to each LTO variant.
LTO variants differ in encoder (Gaussian kernel
vs.\ learned attention).}
\label{tab:darcy-results}
\begin{tabular}{lccccccc}
\toprule
Method & Params & 85 & 106 & 141 & 211 & 421\\
\midrule
LTO (ours) & 29,139 & 0.133 & 0.133 & 0.133 & 0.133 & 0.132  \\
LTO with attn (ours) & 28,894 & 0.055 & 0.055 & 0.055 & 0.055& 0.055  \\
LNO \citep{wang2024latent} & 29,171 & \textbf{0.012} & \textbf{0.011} & \textbf{0.011} & \textbf{0.011} & \textbf{0.011}  \\
\bottomrule
\end{tabular}
\end{table}

\section{Discussion and Limitations}
\label{sec:discussion}

An important feature of the LTO is its direct temporal formulation: given a single source state $u_s$, it maps directly to a requested target time $t$ through $m_{s\to t}$. In the matched comparisons, FNO and LNO receive the same single source state. On Navier--Stokes, FNO and LNO are more accurate at the first step, while LTO gives lower long-horizon errors, with direct evaluation further improving over recursive evolution.

The discretization result controls only the SetConv reconstruction error and assumes $C^2$ states, a periodic domain, and sufficiently well-distributed context points. The autoencoder and latent-evolution errors in \eqref{eq:base-bound} remain separate contributions. The sharp error increase at $128\times128$ on Navier--Stokes further shows that resolution-flexible evaluation does not by itself ensure accurate above-native resolution transfer.

Finally, the SetConv analysis exploits a fixed Gaussian kernel, which restricts context weighting to physical distance. The Darcy experiment illustrates this limitation: replacing the fixed kernel with learned attention substantially reduces the approximation error, but the resulting model is no longer covered by the SetConv analysis developed here. This motivates more adaptive reconstruction operators for which quantitative discretization bounds can still be established.

\section{Conclusion and Future Work} \label{sec:conclusion}

We introduced the Latent Twin Operator, a resolution-flexible encode$\to$evolve$\to$decode surrogate for time-evolving PDEs. For its SetConv realization, we identified the abstract discretization term in the Latent Twin approximation theory with a concrete scattered-data reconstruction error and derived an explicit $\mathcal O(N^{-2/(3D)})$ rate under quasi-uniform refinement. The heat equation experiment supports the predicted decay. On Burgers, LTO achieves lower one-step error than the matched FNO across all tested resolutions. On Navier--Stokes, LTO gives the lowest one-step error at native and sub-native resolutions and lower error at longer horizons, with direct temporal evaluation further reducing error accumulation. The deterioration observed above the native Navier--Stokes resolution, together with the Darcy results, also identifies limitations of the current fixed-resolution latent representation and local SetConv reconstruction.

Future work includes extending the discretization analysis to more general domains and sampling geometries, improving above-native resolution transfer, and developing adaptive or learned reconstruction operators that retain quantitative error control. Further directions include adaptive sensing and experimental design, temporal-consistency constraints for direct latent evolution, and the incorporation of physical or stochastic structure into the latent dynamics.

\subsection*{Reproducibility Statement}

Code for the Latent Twin Operator architecture and all training and
evaluation scripts used in Section~\ref{sec:experiments}, together with
the exact PDEBench data splits for the Burgers and Navier--Stokes
experiments, will be released after acceptance. All
experiments use publicly available PDEBench data
\citep{takamoto2022pdebench}; the FNO baseline for Burgers is trained
from scratch on the same split rather than taken from PDEBench's own
published number, which we could not independently reproduce.
Hyperparameters -- bandwidth $\ell$, latent resolution and channel
count, optimizer and learning-rate schedule, epoch counts -- are
recorded in the released training configs rather than the main text, to
stay within the page limit. For numerical stability, the implementation
adds a small fixed $\varepsilon_{\mathrm{num}}>0$ to the denominator of
the normalized SetConv channel. 

\subsection*{AI Use Statement}

Large language model assistance was used in revising prose and notation in the Introduction, Section~\ref{sec:lto}, Related Work, Discussion, and Experiments sections. All theorem statements, proofs, experimental design choices, and numerical results are the authors' own.

\subsection*{Ethics Statement}
This work concerns surrogate modeling for partial differential equations using synthetic and publicly available benchmark data (PDEBench); it does not involve human subjects, personal data, or any data collection of our own. We do not anticipate ethical concerns specific to this work beyond those generally associated with scientific machine learning research.

\bibliography{iclr2027_conference}
\bibliographystyle{iclr2027_conference}

\appendix
\section{Appendix}
\label{sec:App}
% [TODO] extra ablations, hyperparameters.

\subsection{Deferred Proofs}
\label{app:proofs}

\begin{lem}[Constancy of $\bar\kappa$]\label{lem:kbar-constant}
Under Assumption~\ref{assum:reg} , $\bar\kappa(x) := \int_\Omega \kappa_\ell(x,y)\, \mathrm{d} y$ is constant for all  $x \in \Omega$.
\end{lem}

\paragraph{Proof of Lemma~\ref{lem:kbar-constant} (Constancy of $\bar\kappa$).}
Fix $x\in\Omega$ and substitute $z = y - x \bmod \Omega$ in the integral. Because $\Omega$ is periodic, this substitution is a bijection of $\Omega$ onto itself with Jacobian $1$. Further, since $d_\Omega(x,y) = d_\Omega(0,y-x \bmod \Omega) = d_\Omega(0,z)$ then
$$
\kappa_\ell(x,y) = \exp\big(-d_\Omega(x,y)^2/(2\ell^2)\big) = \exp\big(-d_\Omega(0,z)^2/(2\ell^2)\big) = \kappa_\ell(0,z)\,.
$$
Substituting,
$$
\bar\kappa(x) = \int_\Omega \kappa_\ell(x,y)\,\mathrm{d} y = \int_\Omega \kappa_\ell(0,z)\,\mathrm{d} z = \bar\kappa(0)\,.
$$
Since $x\in\Omega$ was arbitrary, $\bar\kappa(x)=\bar\kappa(0)$ for every $x\in\Omega$.

\begin{lem}[Encoder input stability]\label{lem:encoder-stability}
    For any finite context locations $c_{\ctx}=\{c_1,\dots,c_N\}\subset\Omega$, any bandwidth $\ell>0$, any $\varepsilon>0$, and any two states $u,v\in\calU$,
    $$
    \sup_{x\in\Omega} \big| \hat u_{c_{\ctx}}(x;u) - \hat u_{c_{\ctx}}(x;v) \big| \;\le\; \max_{1\le j \le N} |u(c_j) - v(c_j)| \;\le\; \|u-v\|_\calU\,,
    $$
    i.e.\ $u \mapsto \hat u_{c_{\ctx}}(\cdot\,;u)$ is $1$-Lipschitz, \emph{uniformly over $N$, over $\ell$, over $\varepsilon$, and over the placement of $c_{\ctx}$}.

    By contrast, the raw channel only satisfies
    $$
    \sup_{x\in\Omega} \big|\nu_{c_{\ctx}}(x;u) - \nu_{c_{\ctx}}(x;v)\big| \;\le\; \Big(\sup_{x\in\Omega}\rho_{c_{\ctx}}(x)\Big)\cdot \max_{1\le j\le N}|u(c_j)-v(c_j)|\,,
    $$
    and $\sup_{c_{\ctx}}\sup_{x\in\Omega}\rho_{c_{\ctx}}(x)$ is unbounded as $N\to\infty$ (e.g.\ placing all $N$ context points within an $o(\ell)$ neighborhood of some $x_0$ gives $\rho_{c_{\ctx}}(x_0)\to N$), so no context-count-independent Lipschitz constant exists for the raw channel in general.
\end{lem}
Dividing by the local density turns each context point's contribution into a convex-combination weight, so the normalized channel's output can move by no more than the largest single input perturbation, however many context points there are or however they are placed. The raw channel skips this normalization, so an adversary who clusters points can drive $\rho_{c_{\ctx}}$, and with it the sensitivity, arbitrarily high.

\begin{cor}[Explicit, resolution-independent encoder Lipschitz constant]\label{cor:Le}
    Recall the LTO's encoder $e(c_{\ctx};u) = g_\phi\big(\rho_{c_{\ctx}},\, \hat u_{c_{\ctx}}(\cdot\,;u)\big)$ from Definition~\ref{def:lto}. For a fixed learned network $g_\phi$ (the convolutional stack described above) with Lipschitz constant $L_g$ in its second argument (with respect to $\|\cdot\|_\infty$ on the sampled channel), holding the density channel $\rho_{c_{\ctx}}$ fixed,
    $$
    \|e(c_{\ctx};u) - e(c_{\ctx};v)\| \;\le\; L_g\cdot\|u-v\|_\calU\,,
    $$
    for every context set $c_{\ctx}$, every $N=|c_{\ctx}|$, and every bandwidth $\ell$: the encoder is Lipschitz with a constant that does not depend on the resolution or cardinality of the context set.

    Feeding the raw channel $\nu_{c_{\ctx}}(\cdot\,;u)$ instead of $\hat u_{c_{\ctx}}(\cdot\,;u)$ does \emph{not} admit such a context-count-independent constant, by Lemma~\ref{lem:encoder-stability}: any Lipschitz bound obtained this way would need to grow with $\sup_{x}\rho_{c_{\ctx}}(x)$, and hence, in the worst case, with $N$. The normalized channel is what every experiment in this paper uses.
\end{cor}

\paragraph{Proof of Lemma~\ref{lem:encoder-stability} (Encoder input stability).}
Fix $x\in\Omega$ and write $w_j := \kappa_\ell(x,c_j) \ge 0$, so $\rho_{c_{\ctx}}(x) = \sum_j w_j$ and $\nu_{c_{\ctx}}(x;u) = \sum_j u(c_j) w_j$. Then
$$
\nu_{c_{\ctx}}(x;u) - \nu_{c_{\ctx}}(x;v) = \sum_{j=1}^N \big(u(c_j)-v(c_j)\big)\,w_j\,,
$$
so, since $w_j\ge0$,
$$
\big|\nu_{c_{\ctx}}(x;u) - \nu_{c_{\ctx}}(x;v)\big| \;\le\; \max_j|u(c_j)-v(c_j)|\cdot\sum_j w_j \;=\; \max_j|u(c_j)-v(c_j)|\cdot \rho_{c_{\ctx}}(x)\,. \qquad(\ast)
$$
This proves the raw-channel bound after taking $\sup_{x}$. For the normalized channel,
$$
\hat u_{c_{\ctx}}(x;u) - \hat u_{c_{\ctx}}(x;v) = \frac{\nu_{c_{\ctx}}(x;u)-\nu_{c_{\ctx}}(x;v)}{\rho_{c_{\ctx}}(x)+\varepsilon}\,,
$$
and dividing $(\ast)$ by $\rho_{c_{\ctx}}(x)+\varepsilon$ gives
$$
\big|\hat u_{c_{\ctx}}(x;u)-\hat u_{c_{\ctx}}(x;v)\big| \;\le\; \max_j|u(c_j)-v(c_j)|\cdot\frac{\rho_{c_{\ctx}}(x)}{\rho_{c_{\ctx}}(x)+\varepsilon} \;\le\; \max_j|u(c_j)-v(c_j)|\,,
$$
since $\rho_{c_{\ctx}}(x)/(\rho_{c_{\ctx}}(x)+\varepsilon)\le 1$ for $\rho_{c_{\ctx}}(x)\ge 0,\varepsilon>0$. Taking $\sup_{x\in\Omega}$ and bounding $\max_j|u(c_j)-v(c_j)|\le\|u-v\|_\calU$ gives the claim. The unboundedness of the raw bound follows by placing all $N$ locations within distance $o(\ell)$ of a fixed $x_0$: each $\kappa_\ell(x_0,c_j)\to\kappa_\ell(0,0)=1$, so $\rho_{c_{\ctx}}(x_0)\to N$, which is unbounded as $N\to\infty$. Taking $u-v\equiv1$ realizes this bound directly, $\nu_{c_{\ctx}}(x_0;u)-\nu_{c_{\ctx}}(x_0;v)=\rho_{c_{\ctx}}(x_0)\to N$.

\begin{prop}[Discretization/quadrature error]\label{prop:quadrature}
Under Assumption~\ref{assum:reg}, for every $u\in\calK$ and $x\in\Omega$,
$$
\big|\hat u_{c_{\ctx}}(x;u) - \bar u_\ell(x)\big| \;\lesssim\; C_\calK\cdot\frac{\delta(c_{\ctx})}{\ell}\,,
$$
for $N$ large enough that $(N/|\Omega|)\bar\kappa\gg\varepsilon$, where $C_\calK$ depends on $D$, $\Omega$, the Hessian bound $M$ of Assumption~\ref{assum:reg}, and $\sup_{u\in\calK}\|u\|_\infty$, but not on $N$, $\ell$, or the placement of $c_{\ctx}$.
\end{prop}

\paragraph{Proof of Proposition~\ref{prop:quadrature} (Discretization/quadrature error).}
Fix $x\in\Omega$. Since $|V_j|=|\Omega|/N$ for every $j$,
$$
\rho_{c_{\ctx}}(x) - \frac{N}{|\Omega|}\int_\Omega\kappa_\ell(x,y)\,\mathrm{d} y \;=\; \sum_{j=1}^N \frac{1}{|V_j|}\int_{V_j}\big[\kappa_\ell(x,c_j)-\kappa_\ell(x,y)\big]\,\mathrm{d} y\,.
$$
Taylor-expand $\kappa_\ell(x,\cdot)$ around $c_j$: for $y\in V_j$,
$$
\kappa_\ell(x,c_j)-\kappa_\ell(x,y) \;=\; -\nabla_y\kappa_\ell(x,c_j)\cdot(y-c_j) \;-\; \tfrac12(y-c_j)^\top D_y^2\kappa_\ell(x,\xi)(y-c_j)
$$
for some $\xi$ on the segment $[c_j,y]$. A direct computation gives $D_y^2\kappa_\ell(x,y) = \big(\tfrac{(y-x)(y-x)^\top}{\ell^4}-\tfrac{I}{\ell^2}\big)\kappa_\ell(x,y)$, hence, by the triangle inequality on its two eigenvalue directions, the \emph{pointwise} (not worst-case) operator-norm bound
$$
\|D_y^2\kappa_\ell(x,y)\|_{\rm op} \;\le\; h(y) \;:=\; \Big(\frac{|y-x|^2}{\ell^4}+\frac{1}{\ell^2}\Big)\kappa_\ell(x,y)\,,
$$
which, like $\kappa_\ell$ itself, decays away from $x$; bounding each cell by its own value of $h(c_j)$, rather than a single worst-case constant, is what makes the argument below work in every dimension $D$. Integrating the Taylor expansion above over $y\in V_j$ (diameter $\le2\delta$) and dividing by $|V_j|$, writing $\bar z_j := \frac{1}{|V_j|}\int_{V_j}(y-c_j)\,\mathrm{d}y$ for the cell's average offset from $c_j$ (so $|\bar z_j|\le2\delta$, since every point of $V_j$ lies within $2\delta$ of $c_j$), then summing over $j$,
$$
\rho_{c_{\ctx}}(x) - \frac{N}{|\Omega|}\int_\Omega\kappa_\ell(x,y)\,dy \;=\; -\sum_{j=1}^N \nabla_y\kappa_\ell(x,c_j)\cdot\bar z_j \;-\; E_2\,, \qquad |\bar z_j|\le2\delta\,,
$$
$$
|E_2| \;\le\; 2\delta^2\sum_{j=1}^N h(c_j)\,\big(1+O(\delta/\ell)\big)\,,
$$
the $O(\delta/\ell)$ correction coming from replacing $h(\xi)$, $\xi\in[c_j,y]$, by $h(c_j)$ within each cell, valid since $h$ is itself smooth at the scale $\ell\gg\delta$. Exactly as for the linear term below, $h$ is smooth and decays like $\kappa_\ell$, so the same cell-quadrature estimate applies: $\sum_j|V_j|\,h(c_j) = \int_\Omega h(y)\,\mathrm{d}y\,\big(1+O(\delta/\ell)\big)$. Writing $y-x=\ell w$ (so $dy=\ell^D\,\mathrm{d}w$) gives $\int_\Omega h(y)\,\mathrm{d}y = K_D''\,\ell^{D-2}$, for the finite, $x$- and $c_{\ctx}$-independent constant $K_D'':=\int_{\R^D}(|w|^2+1)e^{-|w|^2/2}\mathrm{d}w$ (periodization error again negligible under Assumption~\ref{assum:reg}). Hence
$$
|E_2| \;\lesssim\; 2\delta^2\cdot\frac{N}{|\Omega|}\,K_d''\,\ell^{D-2}\,.
$$
For the first (linear) term,
$$
\Big|\sum_{j=1}^N \nabla_y\kappa_\ell(x,c_j)\cdot\bar z_j\Big| \;\le\; 2\delta\sum_{j=1}^N \big|\nabla_y\kappa_\ell(x,c_j)\big|\,,
$$
using the \emph{pointwise} gradient at each $c_j$ rather than a coarser cell-wide supremum, for the same reason as the curvature term above: a supremum bound would overcount cells far from $x$. Since $g(y):=|\nabla_y\kappa_\ell(x,y)|=\ell^{-2}|y-x|\,\kappa_\ell(x,y)$ is itself smooth and decays like $\kappa_\ell$, the same cell-averaging step applied to $g$ gives $\sum_j|V_j|\,g(c_j) = \int_\Omega g(y)\,\mathrm{d}y\,\big(1+O(\delta/\ell)\big)$, a standard scattered-data quadrature estimate for a smooth, rapidly-decaying integrand (\citealp{wendland2005scattered}). The integral evaluates exactly, by the same substitution as above, to $\int_\Omega g(y)\,dy = K_d\,\ell^{D-1}$ for the finite, $x$- and $c_{\ctx}$-independent constant $K_d:=\int_{\R^D}|w|e^{-|w|^2/2}\mathrm{d}w$ (periodization error negligible under Assumption~\ref{assum:reg}). Hence
$$
2\delta\sum_{j=1}^N|\nabla_y\kappa_\ell(x,c_j)| \;=\; 2\delta\cdot\frac{N}{|\Omega|}\,K_d\ell^{D-1}\big(1+O(\delta/\ell)\big)\,,
$$
which dominates the bound on $E_2$ above for \emph{every} dimension $d$: comparing the two displays, $E_2$'s contribution is smaller by a factor $\Theta(\delta/\ell)$, with no $D$-dependence left in that ratio, so it suffices that $\ell^\star=\Theta(\delta^{1/3})\gg\delta$ as $\delta\to0$, true in any dimension. Combining,
$$
\Big|\rho_{c_{\ctx}}(x) - \tfrac{N}{|\Omega|}\bar\kappa\Big| \;\lesssim\; \tfrac{N}{|\Omega|}\cdot 2\delta K_d\ell^{D-1}\,,
$$
i.e.\ a \emph{relative} error of order $\delta/\ell$ (using $\bar\kappa\sim\ell^D$). The same argument applies to $\nu_{c_{\ctx}}(x;u) = \sum_j u(c_j)\kappa_\ell(x,c_j)$ against its continuum counterpart $\tfrac{N}{|\Omega|}\bar\kappa\bar u_\ell(x)$, Taylor-expanding the product $u(y)\kappa_\ell(x,y)$ rather than $\kappa_\ell(x,y)$ alone. This introduces one additional term from $\nabla u$, absent above; bounding $|\nabla u|$ via Assumption~\ref{assum:reg}'s Hessian bound $M$ (a standard interpolation inequality on the compact domain $\Omega$) shows this term is $O(\delta\cdot\ell^D)$ after the same cell-quadrature step, strictly smaller than the $O(\delta\cdot\ell^{D-1})$ kernel term as $\ell\to0$, so it does not change the rate. The same relative-error conclusion follows, with an extra factor of $\sup_{u\in\calK}\|u\|_\infty$. Propagating both relative errors through the ratio $m=\nu/(\rho+\varepsilon)$ (negligible $\varepsilon$ for $N$ large) gives the stated bound.

\paragraph{Proof of Theorem~\ref{thm:eps-disc} (Explicit discretization error for the SetConv instantiation).}
Triangle inequality on $\hat u_{c_{\ctx}}(x;u)-u(x) = \big[\hat u_{c_{\ctx}}(x;u)-\bar u_\ell(x)\big] + \big[\bar u_\ell(x)-u(x)\big]$, combining \eqref{eq:bias-bound} and Proposition~\ref{prop:quadrature}, then taking $\sup_{x,u}$ gives the first bound. Minimizing $g(\ell)=C_1\ell^2+C_\calK\delta/\ell$ over $\ell>0$ gives $\ell^\star=(C_\calK\delta/2C_1)^{1/3}$ and $g(\ell^\star)=\Theta(\delta^{2/3})$ by direct substitution.

\paragraph{Proof of the smoothing-bias bound~\eqref{eq:bias-bound}.}
Fix $x\in\Omega$ and $u\in\calK$. For a symmetric, second-order kernel, a Taylor expansion of $u$ around $x$ has its first-order term cancel by the kernel's $z\leftrightarrow-z$ symmetry, leaving only curvature, at rate $\ell^2$ (\citealp{wasserman2006all}, Ch.~5; \citealp{fangijbels1996local}). The usual design-bias term of that classical result vanishes here because $\bar\kappa$ is constant in $x$ by translation invariance (Lemma~\ref{lem:kbar-constant}). Bounding the remaining curvature term by the Hessian bound $M$ from Assumption~\ref{assum:reg} and summing the kernel's second moment over the $D$ spatial dimensions gives, for every $u\in\calK$ and $x\in\Omega$, the stated bound $\sup_{u\in\calK}\sup_{x\in\Omega}\|\bar u_\ell(x)-u(x)\|\le\tfrac12 DM\ell^2$.

\subsection{Extended Related Work}
\label{app:related}

\paragraph{Neural operators.} Learning the solution operator of a family of PDEs directly, rather than solving each instance numerically, has been formalized as operator learning \citep{kovachki2023neural,boulle2024mathematical}. DeepONet \citep{lu2019deeponet} represents the map via a branch/trunk decomposition; the Fourier Neural Operator (FNO; \citealp{li2020fourier}) instead parameterizes it by global convolutions computed via the FFT, which requires the input to lie on (or be resampled onto) a regular grid at both train and test time -- precisely the assumption our construction is designed to avoid.

\paragraph{Implicit neural representations.} A separate line of work represents individual functions or fields directly as coordinate networks, e.g.\ SIREN \citep{sitzmann2020implicit} and Fourier-feature MLPs \citep{tancik2020fourier}; see \citet{essakine2024we} for a survey. These representations are discretization-agnostic for a single field, but do not by themselves give an encode$\to$evolve$\to$decode surrogate for a family of time-evolving states, which is the setting of this paper.

\paragraph{Conditional neural processes and SetConv.} Conditional Neural Processes \citep{garnelo2018conditional} and their convolutional variant, ConvCNP \citep{gordon2020convolutional}, map an arbitrary, variable-size set of observations to a functional representation via a kernel-weighted embedding -- the SetConv primitive underlying our encoder and decoder (Section~\ref{sec:lto}). Our contribution is an explicit discretization-error rate for this construction, not the construction itself.

\paragraph{Latent-space operator learning.} Several recent methods encode a possibly irregular input into a fixed-size latent state, evolve it, and decode back out, demonstrating multi-resolution behavior empirically without formally bounding the resulting error. The Latent Neural Operator \citep{wang2024latent} uses a cross-attention encoder/decoder (Physics-Cross-Attention) that decouples input and output sampling locations, closely paralleling the SetConv construction we analyze here; the Latent Neural PDE Solver \citep{li2025latent} trains an autoencoder onto a coarser latent grid before evolving there; and the Physics-Informed Latent Neural Operator \citep{karumuri2026physics} couples two DeepONets under a PDE-residual training objective rather than paired supervision. Prior work, Latent Twins \citep{chung2026latenttwins} and Physics-conforming Latent Twins \citep{chung2027physics}, introduces the general encode$\to$evolve$\to$decode pattern that the Latent Twin Operator of Definition~\ref{def:lto} instantiates and proves the approximation bound~\eqref{eq:base-bound} motivating this paper, but leaves its discretization-error term $\epsilon_{\disc}(N)$ abstract for any concrete architecture, tackled in this paper.

\paragraph{Broader attention-based operator architectures.} A larger family of Transformer-style operator-learning architectures targets the same problem -- attention over an irregular, potentially large input set -- by two genuinely different means. Transolver \citep{wu2024transolver} and the Latent Spectral Model \citep{wu2023solving} share LNO's idea of compressing the input into a smaller, fixed-size latent representation before mixing it: Transolver learns physics-aware ``slices'' that adaptively pool mesh points into a handful of tokens before attention; the Latent Spectral Model evolves a PDE's solution in a learned, lower-dimensional spectral latent space. GNOT \citep{hao2023gnot}, FactFormer \citep{li2024scalable}, ONO \citep{xiao2023improved}, and the Galerkin Transformer \citep{cao2021choose} instead keep a full or comparably-sized token set -- one token per input point, not a compressed handful -- and make attention over it affordable by other means: GNOT scales cross-attention to large point clouds via heterogeneous, gated normalization; FactFormer factorizes attention along each spatial axis separately; ONO replaces softmax attention with an orthogonal, kernel-based variant; and the Galerkin Transformer replaces softmax normalization with a Petrov-Galerkin-style linear projection, explicitly preserving sequence length rather than reducing it. Closest to our own decoupling of observation and query locations is OFormer \citep{li2022transformer}, which -- like PhCA -- uses cross-attention to map an input point set to a query set at arbitrary locations, though, like GNOT, FactFormer, and the Galerkin Transformer, without compressing to a smaller intermediate representation. None of these seven architectures reports an explicit discretization-error rate in the number of input points; we compare empirically against LNO specifically because it is architecturally closest to the SetConv construction we analyze, not because it is uniquely capable of cross-resolution generalization -- ONO \citep{xiao2023improved} reports the same zero-shot Darcy resolution transfer (train at one resolution, evaluate at another, unseen) that LNO does.

\paragraph{Discretization-invariance critiques.} \citet{gao2025discretization} show empirically that FNO-style operators trained at one resolution need not perform consistently when evaluated at another, terming the resulting gap \emph{discretization mismatch error}, and propose an aliasing-free training pipeline as a fix. \citet{lanthaler2024discretization} instead derive an explicit $O(N^{-s})$ discretization-error rate for FNO on a regular grid, in terms of the input's Sobolev regularity $s$. Theorem~\ref{thm:eps-disc} is the analogous explicit rate for the SetConv/ConvCNP-style construction underlying LNO-type architectures, established for arbitrary, not necessarily gridded, context sets.

\paragraph{How LNO differs from our construction.} Physics-Cross-Attention (PhCA) and our SetConv encoder solve the same problem -- map an arbitrary, variable-size input set to a fixed-size representation, decoupled from the output query locations -- but differ in three ways relevant to this paper's claims. \emph{Mechanism}: PhCA learns query/key/value projections and a softmax attention pattern over $M$ latent positions; our embedding step (the density and signal channels of \eqref{eq:density-signal-channel}--\eqref{eq:normalized-signal-channel}) has no learned parameters at all -- only the CNN $g_\phi$ applied afterward is learned -- so the encoder itself is comparatively lightweight. \emph{Latent structure}: PhCA's $M$ latent positions form an unstructured token set, whereas our latent state $Z\in\mathbb R^{H\times W\times k}$ is a genuine spatial grid, which is what lets the evolution operator $m_{s\to t}$ be a plain, translation-equivariant CNN rather than a further attention or token-mixing block. \emph{Theory}: kernel smoothing has a century of classical nonparametric-statistics theory behind it (Nadaraya--Watson bias, scattered-data quadrature), which is what Theorem~\ref{thm:eps-disc} draws on; no comparable explicit discretization-error rate exists for attention-based encoders. The closest theoretical results, \citet{calvello2025continuum} for attention-based operators and \citet{zhang2025discretization} for a DeepONet-style branch-trunk construction, establish universal approximation but not an explicit rate in the number of input points, leaving this an open question for PhCA-style architectures specifically. \emph{Time evolution}: for time-dependent PDEs, \citet{wang2024latent} decode back to geometric space at every intermediate time step in order to compute the training loss there; our latent flow $m_{s\to t}$ (Section~\ref{sec:lto}) never leaves the latent space between the initial encode and the final decode, chaining as many steps as the rollout curriculum requires without a round trip through geometric coordinates. Section~\ref{sec:exp-ns2d} reports a direct empirical comparison against LNO on a shared 2D Navier--Stokes benchmark, where this architectural difference and the discretization-error guarantee of Theorem~\ref{thm:eps-disc} can be checked against a trained baseline rather than argued from architecture alone.

\subsection{Darcy Flow: Attention mecahsnism and Native LNO Parameter Budget}
\label{app:darcy-native}
Section~\ref{sec:exp-darcy} additionally evaluates a learned-attention
variant of the SetConv encoder. Recall from Section~\ref{sec:lto} that
the fixed Gaussian kernel weights each context point $c_j$ by
$\kappa_\ell(x,c_j) = \exp\!\left(-d_\Omega(x,c_j)^2/(2\ell^2)\right)$,
a function of distance alone. The attention variant instead replaces
this fixed kernel with a learned score,
$$
    \alpha(x,c_j) = \mathrm{softmax}_j\!\left(\frac{q(x)^\top k(c_j)}{\sqrt{d_k}}\right),
$$
where $q,k:\Omega\to\mathbb R^{d_k}$ are learned embeddings of the
query and context positions into a shared $d_k$-dimensional space, and
the $\sqrt{d_k}$ scaling follows standard scaled dot-product attention
\citep{vaswani2017attention}. Unlike $\kappa_\ell$, $\alpha$ is free to
place weight on a distant $c_j$ whenever doing so reduces training
loss.

Finally, Table~\ref{tab:darcy-results} of Section~\ref{sec:exp-darcy} matches
LNO's parameter count to each LTO variant for a controlled comparison.
For reference, Table~\ref{tab:darcy-results-native} reports an
additional comparison using LNO's originally reported
$762{,}113$-parameter configuration. The
resulting ratios ($12.99\times$ for the plain kernel, $3.69\times$
for the attention encoder) are close to the parameter-matched ones,
so the gap in Section~\ref{sec:exp-darcy} is not primarily an
artifact of LNO's larger parameter budget.

\begin{table}[t]
\centering
\caption{Darcy flow: relative $L^2$ error across evaluation
resolutions. LTO variants differ in encoder (Gaussian kernel
vs.\ learned attention).}
\label{tab:darcy-results-native}
\begin{tabular}{lcccccccc}
\toprule
Method & Params & 85 & 106 & 141 & 211 & 421 & Mean Ratio \\
\midrule
LTO (ours)& 29,139  & 0.099 & 0.096 & 0.093 & 0.095 & 0.218 & 12.99$\times$  \\
LTO with attn (ours) & 240,594  & 0.035 & 0.035 & 0.035 & 0.035 & 0.035  & 3.69$\times$  \\
LNO \citep{wang2024latent}    & 762,113 & 0.012 & 0.010 & 0.009 & 0.008 & 0.009  & 1.00$\times$ (ref.)  \\
\bottomrule
\end{tabular}
\end{table}

\end{document}